\documentstyle [12pt,graphics]{article}
\begin{document}

\begin{center}

{\bf Effects of Vacuum Polarization in Strong Magnetic Fields with
an Allowance Made for the Anomalous Magnetic Moments of Particles}
\end{center}

\begin{center}
\textbf{V. N. Rodionov}
\end{center}

\begin{center}
\textit{Moscow State Geological Prospecting University, Moscow, 118873
Russia e-mail: physics@msgpa.ru }
\end{center}
\vspace{2cm}

\begin{abstract}
Given the anomalous magnetic moments of electrons and
posit-rons in the one-loop approximation, we calculate the exact Lagrangian
of an intense constant magnetic field that replaces the Heisenberg-Euler
Lagrangian in traditional quantum electrodynamics (QED). We have established
that the derived generalization of the Lagrangian is real for arbitrary
magnetic fields. In a weak field, the calculated Lagrangian matches the
standard Heisenberg-Euler formula. In extremely strong fields, the field
dependence of the Lagrangian completely disappears, and the Lagrangian tends
to a constant determined by the anomalous magnetic moments of the particles.
\end{abstract}

\vspace{1cm}
\begin{center}
1. INTRODUCTION
\end{center}

The quantum corrections to the Maxwellian Lagrangian of a constant
electromagnetic field were first calculated by Heisenberg and Euler [1] in
1936. The radiative corrections that correspond to the polarization of an
electron-positron vacuum by external electromagnetic fields with diagrams
containing different numbers of electron loops are still the focus of
attention [2-4]. Estimates suggest that the quantum (radiative) corrections
could reach the Maxwellian energy density of the electromagnetic field only
in exponentially strong electromagnetic fields
$F_{c}\sim \exp (‡\pi/\alpha)H_{c}$\footnote{Here, we use a system of units with
$\hbar = c = 1,
H_c =4.41\cdot 10^{13}$ G is the characteristic scale of the electromagnetic field
intensity in QED, $e$ and $m$ are the electron charge and
mass, and $\alpha = 1/137$ is the fine-structure
constant.} [5].
 The calculations by Heisenberg and Euler
are known to contain no approximations in the intensity of external
electromagnetic fields, and their results have been repeatedly confirmed by
calculations performed in terms of different approaches. On this basis,
several authors have identified the field intensity ${F_c}$ with the
validity boundary of universally accepted QED. However, it is clear that,
although such quantities were greatly overestimated, because the
corresponding scale lengths are many orders of magnitude smaller than not
only the scale on which weak interactions manifest themselves, but also the
Planck length, determining the validity range of traditional QED is
currently of fundamental importance. While on the subject of the physics of
extremely small distances, we cannot but say that there is a deep
analogy between the phenomena that arise for large momentum transfers and the processes in intense
electromagnetic fields [2-17]. In fact, the overlapping of the seemingly
distinctly different areas of physics is not accidental and is suggested by
simple dimension considerations.

An allowance for the electromagnetic field intensity based on the exact
integrability of the equations of motion is known to play an important role
in studying the quantum effects of the interaction between charged particles
and the electromagnetic field. In particular, the standard Schwinger
correction to the Bohr magneton
$$
  \mu_{0} = \frac {e} {2m}
$$
which is called the anomalous magnetic moment of a particle,

$$
      \Delta\mu = \mu_{0}\frac{\alpha}{2\pi} ,
$$
manifests itself only in the nonrelativistic limit for weak quasi-static
fields [7]. Indeed, when the influence of an intense external field is
accurately taken into account, the anomalous magnetic moment of a particle
calculated in QED as a one-loop radiative correction decreases with
increasing field intensity and increasing energy of the moving particles
from the Schwinger value to zero. In particular, for magnetic fields
$H \sim  H_c$, the anomalous magnetic moment of an electron is
described by the asymptotic formula [7,10]

 \begin{equation}
\label{eq1}
           \Delta\mu(H) = \mu_{0}\frac{\alpha}{2\pi}\ln\frac{2H}{H_c}
\end{equation}

It follows from (1) that $\Delta\mu(H)$ becomes zero only at one point while
decreasing with increasing field. A similar expression for the anomalous magnetic moment of an electron in an intense
constant crossed field $ {\bf E \perp H}; (E = H)$ at
$H p_{\perp} \gg m H_c$, where $p_{\perp}$ is the electron momentum
component perpendicular to ${\bf[E \times H]}$, can be represented as [11]

 \begin{equation}
\label{eq2}
           \Delta\mu(E) = \mu_{0}
           \frac{\alpha\Gamma (1/3)}{9\sqrt{3}}
           \left(\frac{3 p_{\perp}H}{m H_c}\right)^{-2/3}
\end{equation}
Note that $\Delta\mu(E) \ne 0$ in the entire range of parameters.

Numerous calculations of the Lagrangian for an electromagnetic field (see,
e.g., [1-3, 5,7, 11]) have been performed by assuming that the magnetic
moment of electrons is exactly equal to the Bohr magneton,
 i.e., at
$\Delta\mu = 0\,$.\footnote{ The author of [6] took into account
the anomalous magnetic moments
when analyzing vacuum polarization of arbitrary
spin particles. To this end  the description particles with
anomalous magnetic moments
 in external electromagnetic fields the second order wave equation
was considered.
However,  the  Dirac relativistic wave equations
provide a basis for relativistic quantum mechanics and
quantum electrodynamics of spinor  particles. It is well
known  that the existence
 of a anomalous magnetic moment of an electron can be taken
 into account in the modified Dirac equation
by means of Pauli
interection term $\sim \Delta\mu \sigma^{\mu\nu} F_{\mu\nu}$, were
 $F_{\mu\nu}$--the electromagnetic field strength tensor [8] (see also [7, 18]).}
  However, the following question
is of considerable importance in
elucidating the internal closeness of QED: What effects will an allowance
for the anomalous magnetic moments of electrons and positrons produce when
calculating the polarization of an electron-positron vacuum by intense
electromagnetic fields?

Thus, it is of interest to compare the radiative corrections to the
Maxwelli-an Lagrangian of a constant field calculated by the traditional
method with the results that can be obtained from similar calculations by
taking into account the nonzero anomalous magnetic moments of particles. The
fact that the Lagrangian replacing the Heisenberg-Euler Lagrangian with
nonzero anomalous magnetic moments can be calculated by retaining the method
of exact solutions of the modified Dirac-Pauli equation [8]
in arbitrarily intense
electromagnetic fields also deserves serious attention. In the approach
being developed, the suggested theoretical generalization initially contains
no constraints on the electromagnetic field intensity.

It should be noted that nonzero anomalous magnetic moments also appear in
some of the modified quantum field theories (QFTs) that also describe the
electromagnetic interactions. In particular, this is true for a
generalization of the traditional QFT known as the theory with "fundamental
mass" (see, [12, 13] and references therein). The starting point of this
theory is the condition that the mass spectrum of elementary particles is
limited. This condition can be represented as
 \begin{equation}
m\le M,
 \end{equation}
where the new universal parameter ${M}$ is called the fundamental mass.
Relation (3) is used as an additional fundamental physical principle that
underlies the new QFT. A significant deviation from traditional calculations
is the fact that the charged leptons in QED with fundamental mass have
magnetic moments that are not equal to the Bohr magneton. This is because,
apart from the traditional "minimal" term, the new Lagrangian of the
electromagnetic interaction includes "nonminimal"
terms. Thus, an electron in modified QED has an anomalous magnetic moment
from the outset:
  \begin{equation}
\label{eq4}
           \Delta\mu = \mu - \mu_{0} =
           \mu_{0}\left(\sqrt{1+\frac{m^2}{M^2}}- 1\right)
\end{equation}

An important aspect of the problem under consideration is that the current
state of the art in the development of laser physics [14] allows one to
carry out a number of optical experiments to directly measure the
contributions from the nonlinear vacuum effects predicted by various
generalizations of Maxwellian electrodynamics [15]. Therefore, it should be
emphasized that experimental verification of the nonlinear vacuum effects
with a high accuracy in the presence of relatively weak electromagnetic
fields can also provide valuable information about the validity of QED
predictions at small distances [16, 17]. Note in passing that precision
measurements of various quantities (e.g., the anomalous magnetic moments of
an electron and a muon) at nonrelativistic energies, together with studies of
the particle interaction at high energies, are of current interest in the
same sense.

\begin{center}
2. THE CORRECTION TO THE LAGRANGIAN OF AN ELECTROMAGNETIC FIELD WITH AN
ALLOWANCE MADE FOR THE ANOMALOUS MAGNETIC MOMENTS OF PARTICLES
\end{center}

Let us consider the correction to the Lagrangian of an electromagnetic field
attributable to the polarization of an electron-positron vacuum in the
presence of an arbitrarily strong constant magnetic field by taking into
account the nonzero anomalous magnetic moments of the particles. To solve
this problem, it is convenient, as in the standard approach [5], to
represent the electron-positron vacuum as a system of electrons that fill
"negative" energy levels. For a constant uniform magnetic field, the
Dirac-Pauli equation containing the interaction of a charged lepton with the
field (including the anomalous magnetic moment of the particle) has an exact
solution [18]. In this case, the energy eigenvalues explicitly depend on the
spin orientation with respect to the axis of symmetry specified by the
magnetic field direction. Thus, the energy spectrum of an electron that
moves in an arbitrarily intense constant uniform magnetic field is
  \begin{equation}
\label{eq5}
           E_n (p,H,\zeta) = m\left[
           \frac{p^2}{m^2}+
           \left(\sqrt{|\frac{H}{H_c}|
           (1+2n+\zeta)} + \zeta\frac{\Delta\mu}{2\mu_0}\frac{H}{H_c}
           \right)^2
           \right]^{1/2}
\end{equation}
where ${p}$ is the electron momentum component along the external
field $\bf{ H}$, $n = 0,1,2,...$ is the quantum number of the Landau
levels, and $\zeta = \pm 1$ characterizes the election spin component along the
magnetic field.

 Noting that the radiative correction to the classical density of the
Lagrangian is equal, to within the sign, to the total energy density of the
electron-positron vacuum in the presence of an external field [5]
$$
        {\cal L}' = - W^H.
$$

Let us calculate $W^H$ in a constant magnetic field by taking into account
the anomalous magnetic moment of the electron. Without dwelling on the
details of standard calculations, we represent $W^H$ as
\begin{equation}
 W^{H} = - {\textstyle{{\left| {eH} \right|} \over {\left( {2\pi }
\right)^{2}}}}\int\limits_{ - \infty }^{\infty } {dp\left[ { - \varepsilon
^{ + }_{0} \left( {p} \right) + \sum\limits_{n = 0}^{\infty } {\left[
{\varepsilon ^{ - }{}_{n}\left( {p} \right) + \varepsilon ^{ + }_{n} \left(
{p} \right)} \right]} } \right]} \quad, \label{WH}
\end{equation}
where
\begin{equation}
{\varepsilon}^{\pm}_{n} =
\sqrt {p^2 + m^2
 {\left(
 \sqrt {1 +
2\frac H{H_{c} }n }
 \pm {\frac H{4H^*_{c} } } \right) }^2}. \label{varepsilon}
\end{equation}

Using the Laplace and Fourier integral transforms for the functions that
define (6) and performing summation over Landau levels, we can obtain the
following formula for ${\cal L}'$:
\begin{eqnarray}
{\cal L}' & = & - \frac {m^4 \gamma \,b_1}{8\pi ^2}
\int\limits_{0}^{\infty }
\frac {d\eta }{\eta^2} e^{ - \eta }
\Bigg[  {\rm sh}\left( b \right) \nonumber \\
& + & \frac {1}{\pi }
  \int\limits_{-\infty }^{\infty }
 \frac {e^{-ix} \,dx}{x} \,{\rm ctg}\left(  - i\gamma\, \eta/b_1 + x\gamma
\right){}_{1}F_{2} \left( \{ 1\} ,\{ 1/4,3/4\} ,
- \frac {b^4}{64 x^2} \right)  \Bigg], \nonumber \\ \label{L}
 \end{eqnarray}
where we use the notation
$a_1 = \eta H/2H_{c}^{\ast }$,
 $b_{1} = 1 + \left( {H/4H_{c} ^{\ast }}
\right)^{2}$, \,
  $b = a_1 /b_1$, \,  $\gamma = H/H_{c} $,
  \,
  ${}_{1} F_{2} ( z )$ --is the generalized hypergeometric function.
  Formula (8) is an exact
expression for the Lagrangian with an allowance made for the anomalous
magnetic moment calculated in the one-loop approximation in an arbitrarily
intense magnetic field. An important deviation from the Heisenberg-Euler
Lagrangian is that expression (8) contains the additional field scale
\begin{equation}
    H_{c} ^{\ast } = \frac {m}{4\Delta\mu}, \label{M}
\end{equation}
In the
theory with fundamental mass, it would be natural to call the quantity
\begin{equation}
    H_{c} ^{\ast } = \frac {M^2}{e} = \frac {M^2}{m^2} H_{c}, \label{M}
\end{equation}
a fundamental field.

Passing to the limits of integration over \textit{x} from zero to $\infty$
 in (8)
and using the evenness of the function ${}_{1}F_2 (z)$, we obtain
\begin{eqnarray}
{\cal L}' & = & - \frac {m^4 \gamma \, b_1}{8\pi ^2} \int\limits_{0}^{\infty }
\frac {d\eta }{\eta^2} e^{ - \eta }
\nonumber \\
  \Bigg[  {\rm sh} (b)
& + & \frac 2\pi
  \int\limits_{0}^{\infty }
 \frac {dx}{x} \,
 \frac {{\rm sin}(2\gamma x) \,{\rm cos}(x)+
 {\rm sin}(x)\,{\rm sh}(y)}{{\rm ch}(2y) -{\rm cos}(2\gamma x)} \,\,
 {}_{1}F_{2} ( z) \Bigg] ,\nonumber \\ \label{L1}
  \end{eqnarray}
where $y=\eta\gamma/b_1$, and $z=-b^4/64x^2$.

In particular, it immediately follows from (11) that
$$
  {\rm Im} {\cal L}' = 0
$$

The fact that the Lagrangian ${\cal L}'$ is real for all possible  field
intensities suggests the absence of unstable modes; i.e., the vacuum in a
constant uniform magnetic field in the case under consideration, as in
traditional QED, is stable against the spontaneous production of
electron-positron pairs.

Next, let us separate out the integral over ${x}$ in expression (11).
After several obvious substitutions, it reduces to
\begin{equation}
I = \int\limits_{ 0 }^{\infty } {{du \left [2a_{2}{\rm sin}(u){\rm cos}(b_2 u)
+(1-{a_{2}}^2){\rm sin}(b_2 u) \right ]}
 \over {u \left [1+a_{2}^2 -2a_2 {\rm cos}(u)\right ]}} {}_{1}F_{2}(z_1), \label{I}
\end{equation}
where
$a_2 = e^{-2y}$,   $ b_2 = (2\gamma)^{-1}$,
 $z_1=-{b^4 (2\gamma)^2 \over 64 u^2}$ .
Since the expansions
$$
{{\rm sin}(u) \over {1 + {a_2}^2 -2a_2 {\rm cos}(u)}} =
 {\rm sin}(u)+a_2 {\rm sin}(2u) +{a_2}^2 {\rm sin}(3u)+\cdots;
$$

$$
{ 1-{a_2}^2 \over  {1 + {a_2}^2 -2a_2 {\rm cos}(u)}} =
1+ 2a_2 {\rm cos}(u) +2 {a_2}^2 {\rm cos}(2u)+\cdots   ,
$$
are valid, we obtain for (12)
\begin{equation}
  I = \int\limits_{0}^{\infty}{du \over u}\left [{-{\rm sin}(b_2 u)+
  2\sum\limits_{k=0}^{\infty}{a_2}^k {\rm sin}[u(k+b_2)]}
  \right ]{}_{1}F_{2}(z_1).
  \label{I2}
  \end{equation}

It is easy to see that the following expansion of the function ${}_1 F_2 (z)$
at  zero may be used in a field that is
weak compared to the fundamental field ${H_c}^*$
\begin{equation}
         {}_1 F_2 (z_1) = 1 + {16 \over 3}z_1
          +{256 \over 105}{z_1}^2 + \cdots \label{F}
 \end{equation}
Hence, we obtain for (13)
\begin{equation}
 I = {\pi \over 2}\cdot{{1+a_2}
   \over {1 - a_2}} = {\pi \over 2}\cdot{}{\rm cth}(y), \label{I1}
 \end{equation}
where $y=\eta\gamma$.

Substituting (15) into (11) and performing standard regularization of the
derived integral [5] yields
\begin{equation}
{\cal L}' = - \frac {m^4}{8\pi ^2} \int\limits_{0}^{\infty }
\frac {e^{ - \eta }}{\eta ^3} \,\, \left[ \eta \,\gamma \,
{\rm cth} \left( \eta\, \gamma \right) - 1 - \frac {\eta ^{2}\,\gamma^{2}}{3}
\right] d\eta. \label{L3}
 \end{equation}
Thus, it follows from (16) that in the limit of a weak field, formula (11)
matches the Heisenberg-Euler Lagrangian [1] for an arbitrarily intense
constant uniform magnetic field.

Next, let us consider $H > 4{H_c}^*$. It is easy to verify that in the
limit of extremely strong fields, $H \gg 16{{H_c}^*}^2 /H_c$,
we may again use expansion (14) and can obtain the
following expression for integral (13):
\begin{equation}
I= \frac{\pi}{2}{\rm cth}y,
 \end{equation}
where $y=16\eta {{H_c}^*}^2/H H_c$.
Results (15) and (17) have
a simple meaning: for a sufficiently wide energy gap that separates the
electron and positron states, the terms with large numbers \textit{k} make
the largest contribution to integral (13). However, for magnetic fields
close to the fundamental field, $H \sim  4H^*$, i.e.,
when the gap width
is close to zero, the term with \textit{k} = 0 makes the largest contribution
to the sum in the integrand of (13). In this case, integral (13) can be
calculated exactly. Our calculations yield
\begin{equation}
I=\frac {\pi}2\cdot {\rm ch} \left[ \eta\frac H{2H_c^*}
\frac 1{1+ {(H/4 H_c^*)}^2} \right]. \label{IH}
\end{equation}

The estimates of integral (13) in the three ranges of magnetic fields
($H\ll {H_c}^* , H \sim  4{H_c}*$, and $H\gg {H_c}^*$)
can be represented as a single formula:
\begin{equation}
I = {\pi \over 2}\cdot {\rm ch} \left[ \eta{H \over {2H_c^*}}{1 \over
{1+ {(H/4 H_c^*)}^2}}\right]
{\rm cth}\left[ {\eta\gamma \over {1+{(H/4 H_c^*)}^2}}\right]. \label{OA}
\end{equation}

Substituting (19) into (11) and regularizing the remaining
diverging integral \footnote{First, as usual [5], the part of the integral that does not contain
the magnetic field intensity and that represents the energy of the free
vacuum elections should be discarded. Second, it is necessary to
subtract the contribution proportional to \textit{H}$^{2}$ that has already
been included in the unperturbed field energy. Discarding this term is
related to renormalizing the field intensity and, hence, the charge.
Finally, subtracting a contribution on the order of $H^4/{{H_c}^*}^{4}$
basically corresponds to renonnalizing an additional parameter of the
theory--the anomalous magnetic moment of the particle.}, we obtaine
\begin{equation}
{\cal L}' = - \frac {m^4}{8{\pi}^2} \int\limits_{0}^{\infty}
\frac {d \eta}{{\eta}^3} e^{-\eta}
\left[ \eta \, b_1 \gamma\, \frac{{\rm ch}
 \left[(1+a) \eta \gamma/b_1 \right]}{{\rm sh}
 \left( \eta \gamma/b_1 \right)}
- b_1^2 - \frac {{\eta}^2}{6} {\gamma}^2\left(2+6a+3a^2\right) \right],
\label{L5}
\end{equation}
were $a = H_c/2{H_c}^*$.

\begin{center}
3. ASYMPTOTIC RESULTS
\end{center}

In weak fields ($H\ll H_c$) and in the limit of very
strong magnetic fields ($H\gg 16 {{H_c}^*}^2/H_c $)
the integrand in (20) admits
an expansion into a series. In the first
approximation
$$
  {\cal L}' = \frac{m^4 \gamma^4}{2880\pi^2 {b_1}^2}
  \left[8-15a^2(2+a)^2 \right]\int\limits_{0}^{\infty}d\eta\,\eta\,e^{-\eta},
$$
whence it follows that
\begin{equation}
  {\cal L}' = \frac{m^4 \gamma^4}{2880\pi^2 {b_1}^2}
  \left[8-15a^2(2+a)^2 \right]. \label{L6}
\end{equation}

Thus, the quantum correction to the Maxwellian Lagrangian in the limit of a
weak field ($H\ll H_c$) can be represented as
\begin{equation}
  {\cal L}' =  \frac{m^4}{360 \pi^2}\frac{H^4}{{H_c}^4}
  \left[1-\frac{15}{2}a^2-\frac{15}{2}a^3-\frac{15}{8}a^4
  +{\cal O}(\gamma^2;\gamma^2 a^2) \right],
\end{equation}
where the first term matches the standard Heisenberg-Euler formula. The
first correction to it attributable to the anomalous magnetic moments of the
particles is negative and quadratic in \textit{a.}

In an extremely strong field $H\gg {{16 H_c}^*}^2/H_c$ we
can also obtain from (21)
\begin{equation}
 {\cal L}' =  \frac{m^4}{180 \pi^2 a^4}
 \left(8- 60 a^2 -60 a^3 -15 a^4\right)\left(1-\frac{8}{\gamma^2 a^2} \right). \label{L7}
\end{equation}

According to (23), in the limit of extremely strong fields, the Lagrangian
${\cal L}'$ ceases to depend on the field;
i.e., as the field grows, the quantum correction to the density of the
Lagrangian in the case under consideration asymptotically approaches the
constant
\begin{equation}
 {{\cal L}'}_{\infty} =  \frac {\alpha^2 }{360\pi^2 (\Delta\mu)^4}. \label{const}
\end{equation}

In a sense, the result obtained may be compared with the situation observed
in the standard model, where the cross sections for several processes cease
to increase with energy if, apart from a photon, vector $W^{\pm}-$ and
 $Z^0$-bosons, an
additional diagrams with a Higgs H-boson is included in the analysis. An
allowance made for this diagram reduces the increasing terms in amplitude
and leads to a behavior of the cross sections consistent with the unitary
limit. Since the standard model does not predict the mass of the H-boson,
it may well be that this particle is much heavier than the t-quark, the
heaviest known elementary particle. Thus,  $ M_H \sim  1$ TeV may
prove to be the critical mass that limits the mass spectrum of elementary
particles, i.e., acts as the fundamental mass (see (3))\footnote
{Note in this connection that the central point in the program of
research on the Large Hadron Collider (LHC) at CERN is the search for
Higgs bosons in a mass range up to 1 TeV.}.

By comparing the correction ${\cal L}' $ with the Lagrangian of the Maxwellian
field, we can determine the field intensity
\begin{equation}
  {F_c}^* =\sqrt{\frac{256\alpha}{45\pi}}\frac{{{H_c}^*}^2}{H_c}, \label{f}
\end{equation}
at which $ {\cal L}_0 $ becomes equal to (24). For $H = {F^*}_c$,
the quantum
correction ${\cal L}'$ does not yet reach its asymptotic value of
${{\cal L}'}_{\infty}$.
  A comparison of $ {\cal L}_0 $ and ${\cal L}'$  in other field ranges
clearly shows that the corrections ${\cal L}'$   are always small compared to
the Lagrangian $ {\cal L}_0 $.
 The relative corrections ${\cal L}'/m^4$  derived
from (20) for the anomalous magnetic moments of particles
 $\Delta\mu_1/\mu_0 = 10^{-3}$;
$\Delta\mu_2/\mu_0 = 10^{-3.05}$ and $\Delta\mu_3/\mu_0 = 10^{-3.1}$
 are plotted against magnetic field intensity $\gamma = H/H_c$ in Fig. 1.
\begin{figure}[t]
\includegraphics{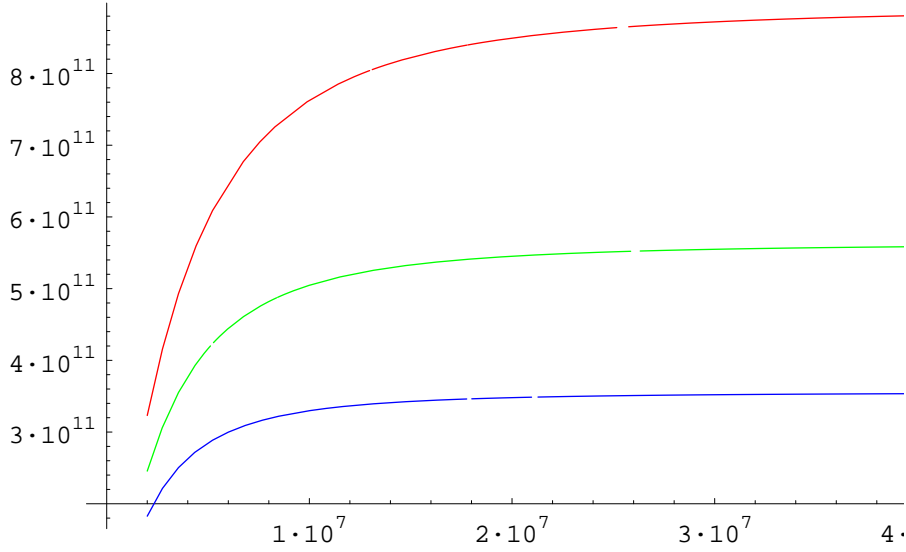}
\caption{Upper curve corresponds to
$\Delta\mu_3/\mu_0=10^{-3.1}$.}
\end{figure}

We estimate the Lagrangian for strong magnetic fields by using formula (20),
which we will represent after the substitution $\gamma\eta/b_1\to x$ as
\begin{equation}
   {\cal L}'= - \frac{m^4 \gamma^2}{8\pi^2}\int\limits_{0}^{\infty}
   \frac{e^{-b_1 x/\gamma}}{x^3}\left[ x \frac {{\rm ch} [x(1+a)]}{{\rm sh}(x)}
   -1 -\frac {x^2}{3}\left(1 + 3 a +\frac{3}{2} a^2 \right)\right] dx.
   \label{L8}
\end{equation}
For $H_c << H << 16{{H_c}^*}^2/H_c$ the range $1<< x << 16{{H_c}^*}^2/{H_c}^2$
is important in integral (26). In this
case, the hyperbolic functions may be substituted with exponentials, and the
integrand in (26) becomes
\begin{equation}
  \frac {e^{f_1(a,\gamma, x)}}{x^2} -\frac{e^{f_2(a,\gamma, x)}}{x^3}
  -\frac{e^{f_2(a,\gamma, x)}}{3 x}\left(1 + 3 a +\frac{3}{2}a^2 \right),
  \label{f}
\end{equation}
where
$$
  f_1 (a,\gamma, x)=-\frac{1}{4\gamma}(2 - a\gamma)^2 x
$$
¨
$$
  f_2 (a,\gamma, x)=-\frac{1+\frac{a^2}{4}\gamma^2}{\gamma} x.
$$

For $H_c << H < 4 {H_c}^*\,\,$ ($ 1 << \gamma < 2/a)\,\, $
$f_1 \sim f_2 = - x/\gamma$ and we find
from (26) with a logarithmic accuracy that
\begin{equation}
    {\cal L}'=\frac{m^4 \gamma^2}{24\pi^2}
    \left(1 + 3 a +\frac{3}{2}a^2 \right){\rm ln}\gamma,\label{GE}
\end{equation}

For $a\to 0$ this formula matches the Heisenberg-Euler Lagrangian
in the limit of strong magnetic fields, $H>>H_c$ [5].

If  $4{H_c}^* < H << 16 {{H_c}^*}^2/H_c $ or
$2/a < \gamma <<  4 / a^2  $, then $f_1 \sim f_2 = -\gamma a^2 x/4$,
and the range $1 << x <<4/(a^2 \gamma)  $ gives the
largest contribution to the integral. In this case, we find from (26) that
\begin{equation}
    {\cal L}'=\frac{m^4 \gamma^2}{24\pi^2}
    \left(1 + 3 a +\frac{3}{2}a^2 \right) \left(2{\rm ln}\frac{2}{a}
    - {\rm ln}\gamma\right). \label{our}
\end{equation}
For $H >> 16 {{H_c}^*}^2 /H_c$, we return to the cases considered
above (see (23)) where the range $x<< 1 $ gives the largest
contribution to integral (26).

If $a\gamma = 2$, i.e., for $H = 4H_c$ the exponent $f_1$, in one of
the exponentials in (27) becomes zero. It is easy to verify that this term
is attributable to the contribution from the ground energy state
$\varepsilon_0$ (see
formula (7)) in which the dependence on particle mass completely drops out at
this field intensity. There is no such state with a "dropping" mass in the
structure of the Heisenberg-Euler Lagrangian for a fixed field. However, if
we consider the passage to the limit $m^{2} \to 0$, then the
Heisenberg-Euler Lagrangian can simulate such an effect. It is easy to see
that the total contribution of the ground state is small compared to the
contribution of the last term in (27), which owes its origin to the field
renormalization in expression (11). A similar conclusion can also be reached
by considering integral (16).

The following should be emphasized when commenting the analogy between the
limits $m^{2} \to 0$ in the Heisenberg-Euler Lagrangian (see formula (16))
and $H \to 4{H_c}^*$ in (26). As we showed above, for $H = 4{H_c}^*$
in the modified Lagrangian, just as in the Heisenberg-Euler Lagrangian
for $m^{2} = 0$, the exponent in the terms whose contribution is vanishingly
small against the background of the contributions from the renormalization
procedure becomes zero. In other words, in both cases, the ground states in
the structure of the integrand are equally preferential, but their
contribution to the integral is not dominant.

Neglecting the first and the second terms in (27), we find from (26) that
$$
    {\cal L}'=\frac{m^4 }{6\pi^2 a^2}
    \left(1 + 3 a +3a^2/2\right){\rm ln}\left(\frac{2}{a}\right).
$$
This result agrees with formulas (28) and (29) from which it can be obtained
by the substitution $\gamma = 2/a$.
Thus, these functions are continuously joined
at $H = 4{H_c}^*$.

Finally, let us estimate the Lagrangian ${\cal L}'$ in terms of traditional
QED, i.e., by taking into account the nonzero anomalous magnetic moments of
particles in intense electromagnetic fields attributable to radiative
effects. Substituting $\Delta\mu $ from (1) into the expression
$b_1 = 1 + (\Delta\mu H/m)^2$ yields an estimate of
${\cal L}'$  (26) in the limit of
extremely strong fields. For a constant magnetic field,
$\gamma>> \mu_0/\Delta\mu$
$$
          b_1 \sim \alpha_2{\rm ln}^2 (2\gamma),
$$
where
$$
        \alpha_2 = \alpha^2/16\pi^2.
$$
In this case, the exponent in (26) is
$$
        f_3 = -x\frac{b_1}{\gamma} = -\frac{\alpha_2{{\rm ln}}^2 (2\gamma)}{\gamma} x.
$$
If $\gamma >> \alpha_2 {{\rm ln}}^2  (2\gamma)$,
 then the range $1 << x << \gamma/\alpha_2{\rm ln}^2 (2\gamma)$
  is important in
(26). Thus, we can find from (26) that
\begin{equation}
    {\cal L}'=\frac{m^4 \gamma^2}{24 \pi^2}
    \left[{\rm ln}\gamma -{\rm ln}\alpha_2-2{\rm ln}({\rm ln}2\gamma)\right]. \label{30}
\end{equation}
The first term in (30) is identical to its estimate in the Heisenberg-Euler
theory [5]. The relative effective Lagrangian ${\cal L}'/m^4$ derived
from the integral representation (26) (with $\alpha_2 \sim  3.4 \cdot 10^{7}$)
is plotted against magnetic field intensity in Fig. 2 (curve
$1$). For comparison, the same figure also shows a plot for
$\Delta\mu = 0$ (curve $2$).
\begin{figure}[t]
\includegraphics{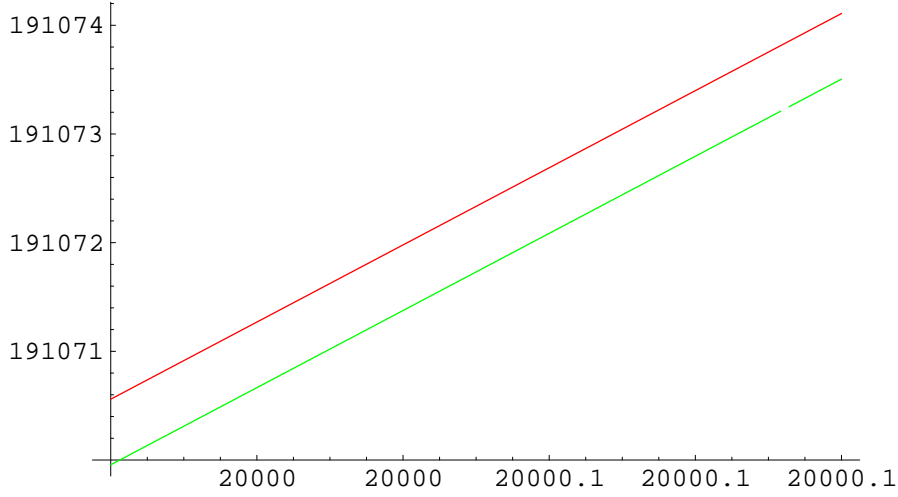}
\caption{Upper curve corresponds to $\Delta\mu=0$.}
\end{figure}

By similar urguments
in the case of an intense constant crossed field
$$
\gamma^{2/3}\alpha_1 >>1,
$$
were
$$
\alpha_1 =
       \frac{\alpha^2 \Gamma^2(1/3)}{972}
      \left(\frac{3 p_{\perp}}{m}  \right)^{-4/3},
$$
we have
$$
      b_1 \sim \gamma^{2/3} \alpha_1.
$$
For exponent  (\ref{L8}) in this case, we fined
$$
        f_3 = -x\frac{b_1}{\gamma} = -\frac{\alpha_1}{\gamma^{1/3}}x.
$$
If $\gamma^{1/3} >> \alpha_1 $, then the range
$1<< x << \gamma^{1/3}/\alpha_1 $ is important in
(\ref{L8}).
 Thus, we can find from (\ref{L8})
\begin{equation}
   {\cal L}' = \frac{m^4 \gamma^2}{72\pi^2}
    \left( {\rm ln}\gamma -3{\rm ln}\alpha_1\right).  \label{100}
\end{equation}

It follows from (\ref{100}), that in intense crosed
field the presence of $\Delta\mu(E) \ne 0$ decreases to  three times
the radiative correction ${\cal L}'$ then analogous result
under the condition  $\Delta\mu(E) = 0$
$$
  \frac{{\cal L}'}{{\cal L}_0} = \frac{\alpha}{9\pi} {\rm ln}\gamma.
$$
The relative effective Lagrangian ${\cal L}'/m^4 $
at $m/p_{\perp} = 0.2$ (with $\alpha_1 \sim 10^{-8}$)
is plotted against crossed field
intensity in Fig.3 (curve $I$).
For comparison, the same figure
shows a plot  for $\Delta\mu = 0$ (curve $2$).
\begin{figure}[t]
\includegraphics{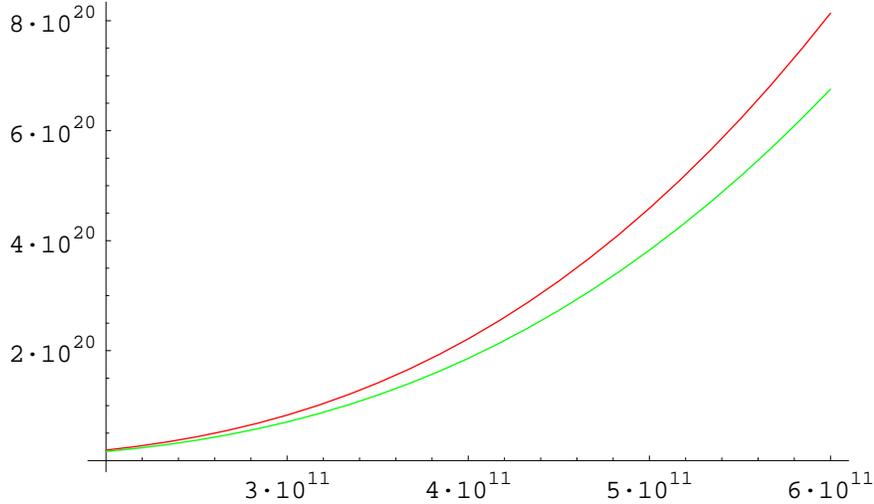}
\caption{Upper curve corresponds to $\Delta\mu=0$.}
\end{figure}

Note that an allowance made for the anomalous magnetic moments of vacuum
particles in terms of universally accepted QED leads to a decrease in the
radiative correction to the field energy density. Recall that we reached a
similar conclusion by considering the static anomalous magnetic moment that
arises, in particular, in the modified field theory. Thus, irrespective of
the nature of the anomalous magnetic moment attributable to the dynamic or
static types of interaction, we obtain consistent results. Our conclusions
are also important in studying the anomalous magnetic moment as the most
accurate calculable and measurable (in
numerous precision experiments) characteristic of particles.

\begin{center}
4. CONCLUSIONS
\end{center}

Our results can be of considerable importance in constructing astrophysical
models, in particular, in  studying extremely magnetized neutron
stars--magnetars; interest in the existence of the latter objects has
increased appreciably in recent years (see, e.g., [4] and references
therein). According to models for the macroscopic magnetization of bodies
composed mostly of neutrons, the intensity of the magnetic fields frozen
into them increases from the surface to the central regions and can reach
$10^{15} -10^{17}$ G [19].

Note also that the radiative effects can be enhanced by external intense
electromagnetic fields not only in Abelian, but also in non-Abelian quantum
field theories. For example, an allowance made for the influence of an
external field on such parameters as the lepton mass and magnetic moment in
terms of the standard model leads to nontrivial results. In this case, apart
from the electrodynamic contribution, the one-loop mass operator of a
charged lepton also contains the contributions from the interaction of
 $W^\pm$-, $Z^0$-, and H-bosons with a vacuum. It is easy to see that,
  in the absence of
an external field, the contribution from weak interactions to the radiative
shift of the lepton mass \textit{m} is suppressed by a factor of
${(m/M_i)}^{2}$ ($i = W, Z, H$) compared to the
electrodynamic contribution. However, the contributions of weak currents in
the ultrarelativistic limit can dominate in intense external fields, as was
first noted in [20] (see also [21]).

In close analogy with the quantum corrections to the particle masses, the
anomalous magnetic moments of charged leptons in the standard model are
attributable to the vacuum radiative effects of electromagnetic and weak
interactions and contain the contribution from the hadron polarization of
the vacuum. For example, for the anomalous magnetic moment of a muon,
$$
   a_{\mu}^{SM} = a_{\mu}(QED) + a_{\mu}(weak) + a_{\mu}^{had}.
$$

According to recent theoretical estimates made in the standard model [22],
the contributions from electromagnetic and weak interactions can be written
as
 $$
      a_{\mu}(QED) = 11 658 470.57(0.29)\cdot 10^{-10};
      $$
      $$
      a_{\mu}(weak) = 15.1(0.4)\cdot 10^{-10}.
      $$

Although the calculations of the contributions from the
hadron polarization of a vacuum to ${a_\mu}^{SM}$ have a history
that spans almost forty years, ${a_\mu}^{had}$ is currently known
with the largest uncertainty (see, e.g., [23-28]). One of the most reliable
estimates for the contributions of the
lowest-order hadron polarization of a vacuum that generalizes the data on
hadron t-decay and $e^+ e^-$ annihilation appears as follows [23,24]:
\footnote{See, however, [25], where the contribution of the highest orders of
hadron polarization of a vacuum were calculated, and the recent papers [26,
27], in which the contribution from the third-had order diagrams to
${a_\mu}^{had}$ attributable to photon-photon scattering was taken into
account.}
$$
    {a_\mu}^{had}  = 692(6)\cdot 10^{-10}
$$
The theoretical anomalous magnetic moment of a muon in the standard model
takes the form [28]
$$
          {a_\mu}^{SM} = 11659177(7)\cdot 10^{-10}
$$

The results of one of the most resent $(g -2)$ experiments aimed at
measuring the anomalous magnetic moments of positive polarized muons carried
out on a  storage ring with superconducting magnets at Brookhaven National
Laboratory (BNL) can be represented as
\begin{equation}
{a_\mu}^ {exp} = 11659204(7)(5)\cdot 10^{-10}.
\end{equation}

The data obtained yield the
difference
\begin{equation}
\Delta_\mu =  {a_\mu}^ {exp} - {a_\mu}^{SM} = 27 \cdot 10^{-10}
\end{equation}
which exceeds the total measurement errors and the uncertainties of the
theoretical estimates. According to the most recent reports from the BNL
muon $(g - 2)$ collaboration [29], the relative value of this excess
is 2.6. A twofold increase in this accuracy is expected in the immediate
future. Clearly, the solution of the muon $(g - 2)$ problem may lead
to the appearance of a new theory outside the scope of the standard theory.

Recall in this connection that the anomalous magnetic moment of a muon in
the modified theory contains the contribution attributable to the new
universal parameter \textit{M} from the outset. According to (4),
\begin{equation}
 a_{\mu}(M)= \frac { {m_{\mu}}^2 }{2 M^2},  \label{a}
\end{equation}
where $\,\,$ $m_{\mu}$ -- is the muon mass.
It is easy to see that $a_{\mu}(M)$
is equal in order of magnitude to (32) at $M \sim 1 $  TeV.

The principal conclusion drawn from a comparison of the above estimates is
that we cannot rule out the possibility that the observed difference between
the theoretical and experimental values for $\Delta_\mu $
is equal to $a_\mu (M)$.
As was pointed out above, the parameter $ M$ in the new theory may be
related to the Higgs boson mass ($M_H$). In this case, tile
difference between ${a_\mu}^ {exp}  $ and ${a_\mu}^ {SM}$  can provide valuable
information about the particle whose mass has not been determined in the
standard model. Substituting $m_{\mu}$ and the anomalous magnetic
moment of a miion into (33), we can easily impose the following
constants on the $H$-boson mass:
$$
           1.2 \, {\rm TeV}  \le M_H \le 1.8 \, {\rm TeV}.
$$

The standard model with the Higgs boson mass $M_H \ge 1$ TeV entails
several additional features, in particular, the impossibility to describe
the weak interactions in the sector of $H, W,$ and $Z$-particles in
terms of the perturbation theory [30]. Naturally, the need for constructing
a new nonperturbadve theory arises in this case. Apart from the condition
for the mass spectrum being limited, $m \le M$ (see (2)),
the Higgs mechanisms of mass formation and compensation for the
discrepancies can become integral elements of one of the most promising
versions of the modified theory--the standard model with fundamental mass.
\vspace{2cm}
\begin{center}
ACKNOWLEDGMENTS
\end{center}

I am grateful to V.G. Kadyshevsky for helpful discussions and valuable
remarks. This study was supported by the Russian Foundation for Basic
Research (project no. 02-02-16784). I am also thanks for partial
support Grant of President of RF (project no. Scientific Schools-2027.2003.2).

\end{document}